\documentclass[twocolumn]{pasj00}

\def\la{\lambda}

\def\D{\Delta}

\usepackage[dvips]{color}

\begin{document}
\SetRunningHead{}{}
\Received{2010/09/06}
\Accepted{}

\title{
A Possible Tilted Orbit of the Super-Neptune HAT-P-11b$^*$}

\author{
Teruyuki \textsc{Hirano},\altaffilmark{1,2}
Norio \textsc{Narita},\altaffilmark{3}
Avi \textsc{Shporer},\altaffilmark{4,5}\\
Bun'ei \textsc{Sato},\altaffilmark{6}
Wako \textsc{Aoki},\altaffilmark{3}
and 
Motohide \textsc{Tamura}\altaffilmark{3}
}

\altaffiltext{1}{
Department of Physics, The University of Tokyo, Tokyo, 113-0033, Japan
}
\email{hirano@utap.phys.s.u-tokyo.ac.jp}
\altaffiltext{2}{
Department of Physics, and Kavli Institute for Astrophysics and Space Research,\\
Massachusetts Institute of Technology, Cambridge, MA 02139, USA
}

\altaffiltext{3}{
National Astronomical Observatory of Japan, 2-21-1 Osawa,
Mitaka, Tokyo, 181-8588, Japan
}

\altaffiltext{4}{
Department of Physics, Broida Hall, University of California, Santa Barbara, CA 93106, USA
}

\altaffiltext{5}{
Las Cumbres Observatory Global Telescope Network, 6740 Cortona Drive, Suite 102, 
Santa Barbara, CA 93117, USA
}

\altaffiltext{6}{
Department of Earth and Planetary Sciences, Tokyo Institute of Technology,
2-12-1 Ookayama, Meguro-ku, Tokyo 152-8551
}


\KeyWords{
stars: planetary systems: individual (HAT-P-11) ---
stars: rotation --- 
techniques: radial velocities --- 
techniques: spectroscopic ---
techniques: photometric}

\maketitle

\begin{abstract}
We report the detection of the Rossiter-McLaughlin effect for
the eccentric, super-Neptune exoplanet HAT-P-11b,
based on radial velocity measurements taken with HDS, mounted on the Subaru 8.2m 
telescope, and 
simultaneous photometry with the FTN 2.0m telescope,
both located in Hawai'i.
The observed radial velocities during a planetary transit of HAT-P-11b
show a persistent blue-shift, suggesting a spin-orbit misalignment
in the system.
The best-fit value for the projected spin-orbit misalignment angle is 
$\la= 103_{~-19^\circ}^{\circ+23^\circ}$.
Our result supports the notion that eccentric exoplanetary systems
are likely to have significant spin-orbit misalignment
(e.g., HD~80606, WASP-8, WASP-14, WASP-17, and XO-3).
This fact suggests that not only hot-Jupiters but
also super-Neptunes like HAT-P-11b had once
experienced dynamical processes such as planet-planet scattering
or the Kozai migration. 
\end{abstract}
\footnotetext[*]{Based on data collected at Subaru Telescope,
which is operated by the National Astronomical Observatory of Japan.}

\section{Introduction\label{s:sec1}}
Transiting exoplanetary systems provide us a unique probe to investigate the
dynamical history of planetary systems discovered today. The Rossiter-McLaughlin 
effect (hereafter, the RM effect), 
which was originally discussed for stellar eclipsing binaries
\citep{Rossiter1924, McLaughlin1924}, 
is an apparent radial velocity (hereafter, RV) anomaly during a planetary transit 
caused by a partial occultation of the rotating stellar surface 
(e.g., \cite{Queloz2000, Winn2005, Ohta2005, Gaudi2007}). 
Through the RM effect, one can estimate the angle between the stellar rotation axis
and the planetary orbital axis projected onto the sky plane.
This angle, which we denote by $\la$, is strongly related to formation and migration
history of close-in exoplanets.

The standard formation theory of close-in gas-giants (hot-Jupiters) suggests
they formed outside the so called ``snow-line'', which is usually located at a few AU away 
from the host star, and then migrated inward due to some migration process
(e.g., \cite{Lin1996, Chambers2009, Lubow2010}).
While migration processes such as disk-planet interactions (type I or II migration) 
predict relatively small values of $\la$, 
dynamical processes including planet-planet scattering
and Kozai cycles might produce a large value of $\la$ (e.g.,
\cite{Wu2007, 2007ApJ...669.1298F, Nagasawa2008, Chatterjee2008}).
The observed distribution of $\la$ and its host star dependence
would help reveal the dynamical history of exoplanetary systems.

To this point, the RM effect has been measured for approximately 30 transiting hot-Jupiters
(see Table~1 of \cite{Winn2010}). 
The observational results indicate that the spin-orbit axes in some of the systems
are well-aligned (at least on the sky), but others 
show significant misalignment. Eccentric exoplanetary systems, where $e\gtrsim 0.1$, 
and systems whose central stars are massive ($M_\star \gtrsim 1.2 M_\odot$)
are more likely to show strong misalignment \citep{Winn2010}. 
Specifically, out of the thirteen eccentric transiting systems 
(CoRoT-9, CoRoT-10, GJ~436, HAT-P-2, HAT-P-11,
HAT-P-14, HAT-P-15, HD~17156, HD~80606, WASP-8, WASP-14, WASP-17, and XO-3), 
the RM effect has been observed for seven systems (HAT-P-2, HD~17156, HD~80606, WASP-8, 
WASP-14, WASP-17, and XO-3). 
Among them, five systems have been reported to have significant spin-orbit misalignment
(HD~80606, WASP-8, WASP-14, WASP-17, and XO-3)
based on measurements of the RM effect \citep{2008A&A...488..763H, Winn2009b, Queloz2010, Winn2009, Triaud2010}.
In addition, \citet{Schlaufman2010} reported the possibility of spin-orbit 
``misalignment along the line-of-sight''
in other eccentric systems (e.g., HD~17156 and HAT-P-14) based on an
analysis of stellar rotational periods (note that the sky-projected
spin-orbit alignment angle for HD~17156 has been reported
as $\la= 10.0^{\circ} \pm 5.1^{\circ}$ by \cite{Narita2009a}).

In this paper, we focus on the RM effect of the super-Neptune HAT-P-11b.
So far, the RM effect has been observed only for hot-Jupiters.
In order to obtain a clearer insight into planetary migration processes,
we need to measure the RM effect for 
a wider range of planetary and stellar parameters.
The transiting super-Neptune exoplanet HAT-P-11b \citep{Bakos2010} was detected
by the HATNet transit survey \citep{Bakos2004} and confirmed by subsequent RV measurements at Keck with HIRES.
HAT-P-11b orbits a bright K dwarf star ($V\sim 9.6$ mag) with an orbital period of
$\sim 4.9$ days. 
Its planetary radius, $R_p=0.422\pm 0.014~R_J$, is one of the smallest among the known transiting exoplanets. 
Although HAT-P-11b is a difficult target for an RM measurement due to its small size, 
the significant eccentricity ($e=0.198\pm 0.046$) makes it very interesting for attempting the 
challenging observation.
The detection of the RM effect for Neptune-sized planets are of great importance for making 
further progress in studying migration histories.

The rest of the present paper is organized as follows.
In Section \ref{s:sec2} we report on new spectroscopic and photometric observations
of the HAT-P-11 system using the High Dispersion Spectrograph (HDS) installed 
on the 8.2m Subaru Telescope, and
the LCOGT 2.0m Faulkes Telescope North (FTN).
The new observations include simultaneous transit spectroscopy and photometry
of HAT-P-11b on UT 2010 May 27 as well as several out-transit RV datasets to determine the orbital 
(Keplerian) motion of HAT-P-11. Data analysis is presented in Section \ref{s:sec3}. 
We combine the new photometric and RV dataset with the published RV data by \citet{Bakos2010} 
and simultaneously determine the orbital and RM parameters in Section \ref{s:sec4}.
We report that the RV anomaly during the transit shows a possible spin-orbit misalignment
in the system.
Finally, Section \ref{s:sec5} summarizes our findings.

\section{Observations\label{s:sec2}}

\subsection{Subaru Spectroscopy}
We conducted spectroscopic observations of HAT-P-11 with Subaru/HDS
on UT 2010 May 21, 27, and July 1. We employed the Std-I2b setup on May 21 and 27, and
the Std-I2a setup on July 1. 
Due to the bad seeing during the May observations,
we needed to broaden the slit width, 
which we set to 0.8$^{\prime\prime}$, yielding a spectral resolution of $R\sim 45000$.
We set it to 0.4$^{\prime\prime}$
for the July observation, corresponding to $R\sim 90000$.
We observed the target with the Iodine cell for a precise wavelength calibration. 
Adopting exposure times of $360$-$420$ seconds, 
we obtained a typical signal-to-noise ratio (S/N) of 150--200.

We reduced the raw data with the standard IRAF procedure, and extracted
one-dimensional spectra. 
We then input the spectra into the RV analysis routine. The RV analysis for Subaru/HDS is 
described in detail by \citet{Sato2002} and \citet{Narita2007}.
Specifically, in order to obtain the stellar template spectrum,
we adopted the method developed by \citet{Butler1996}, which uses a
high S/N, high resolution observed spectrum of the host star without the Iodine cell.
We took that stellar template spectrum during the July~1 observation, and deconvolve it
with the instrumental profiles, which were estimated by the rapid rotator plus I$_2$ spectrum.
The output relative RVs are summarized in Table~\ref{hyo1}.
The reported errors based on the RV analysis do not include stellar ``jitter'', which have been
reported to be significant for HAT-P-11.
The measured RVs as well as the published Keck data by \citet{Bakos2010} are plotted in Figure \ref{RV_all}.

\begin{table}[htb]
\caption{Radial velocities measured with Subaru/HDS.}\label{hyo1}
\begin{center}
\begin{tabular}{lcc}
\hline
Time [BJD (TDB)]  & Relative RV [m~s$^{-1}$] & Error [m~s$^{-1}$]\\
\hline\hline
2455338.06263& 21.81& 3.50\\
2455338.06814& 25.54& 3.17\\
2455343.88425& 2.48& 2.61\\
2455343.89527& 4.69 &2.27\\
2455343.90078& -0.64 & 2.53\\
2455343.90629& 8.59 &2.73\\
2455343.91180& -1.56& 2.52\\
2455343.91731& -3.65& 2.24\\
2455343.92282& 3.52& 2.46\\
2455343.92833& -1.20& 2.45\\
2455343.93384& -4.11& 2.20\\
2455343.93935& -0.79& 2.55\\
2455343.94487& -2.32& 2.27\\
2455343.95038& -2.89& 2.29\\
2455343.95590& -0.98& 2.50\\
2455343.96142& -0.79& 2.48\\
2455343.96694& -1.78& 2.60\\
2455343.97245& 0.94 &2.67\\
2455343.97797& 1.69 &2.29\\
2455343.98349& 1.83 &2.58\\
2455343.98901& 1.82 &2.48\\
2455343.99453& -2.05& 2.43\\
2455344.00004& 4.62 &2.02\\
2455344.00556& 0.70 &2.80\\
2455344.01108&-4.27& 2.72\\
2455344.01659&-4.03& 2.62\\
2455344.02211&-5.10& 2.32\\
2455344.02762&0.28 &2.76\\
2455378.96572&-11.69& 3.45\\
2455378.97054&-8.31 &3.74\\
2455378.97553&-8.92 &3.75\\
2455378.98035&-3.12 &3.54\\
2455378.98517&-4.25 &3.76\\
2455379.09717&-5.42 &3.77\\
2455379.10198&-4.86 &3.62\\
2455379.10683&-4.84 &3.73\\
2455379.11165&2.36 &4.05\\
2455379.11648&-6.53& 3.29\\
2455379.12133&-5.33& 3.93\\
2455379.12615&1.63 &3.55\\

\hline
\end{tabular}
\end{center}
\end{table}

\subsection{Simultaneous FTN Photometry}
In order to derive an accurate estimate of the start and end times of the 
transit, we obtained a photometric light curve of the same 
transit event on UT 2010 May 27. We observed with FTN and the 
Spectral Instruments camera with the Pan-STARRS-Z filter. The camera has a back-illuminated 
Fairchild Imaging CCD and we used the default 2 $\times$ 2 pixel binning mode, with an effective 
pixel scale of 0.304 arcsec pixel$^{-1}$. The telescope was defocused and the 10.5 $\times$ 10.5 
arcmin field of view (FOV) was positioned so the guiding camera FOV will contain a suitable 
guide star. We used an exposure time of 10 seconds and the median cycle time was 30.6 
seconds. FTN observations started at UT 2010 May 27 09:08 and ended at UT 2010
May 27 13:09, an overall duration of 4.0 hours. 
Photometry was done with aperture photometry and the light curve was calibrated 
using several non-variable stars in the field.
Observing conditions have deteriorated at FTN during the 
HAT-P-11 transit observation, and only a small amount of sufficient quality exposures were 
obtained during egress. 
The dome had to be closed soon after the transit ended. 
The resultant photometric light curve is shown in Figure \ref{RV_transit} (panel b).

\section{Analysis\label{s:sec3}}
In order to estimate the positions of the planet on the stellar surface
based on the observed 
RM velocity anomaly during the transit, one must know the relation between
them. Following the procedure described by \citet{Winn2005} and \citet{Hirano2010},
we performed a mock data simulation, by changing the relative decrease in flux $f$ (which is the flux ratio
of the occulted portion on the stellar surface to the disk-integrated flux)
and the velocity component
$v_p$ (which is a line-of-sight component of the rotational velocity of the occulted stellar portion),
we generated many simulated spectra during a transit, and put them into the RV analysis procedure
as we did for the observed spectra.
After fitting the output velocity anomaly $\D v$ with input parameters $f$ and $v_p$ 
we derived the following empirical relation:
\begin{eqnarray}
\D v = -f v_p \left[1.16- 0.205\left(\frac{v_p}{v \sin i_s}\right)^2\right], \label{eq3.1}
\end{eqnarray}
where $v\sin i_s$ is the projected stellar rotational velocity of HAT-P-11, which we 
set here as $v \sin i_s=1.5$ km s$^{-1}$\citep{Bakos2010}.

\citet{Bakos2010} reported that the HAT-P-11 system shows a long-term RV drift 
over one-year of observations of the system. 
Since our new observations span only about one month
we could not determine the long-term RV drift from our new Subaru dataset alone.
Instead, we assumed the RV drift as $\dot{\gamma}=0.0297\pm 0.0050$ m s$^{-1}$ day$^{-1}$,
as reported by \citet{Bakos2010}. 

Our single transit light curve could not refine the light curve parameters 
determined by \citet{Bakos2010}, based on several transit light curves.
We thus adopt the system parameters of \citet{Bakos2010} for 
the semi-major axis scaled by the stellar radius, $a/R_s=15.58^{+0.17}_{-0.82}$, 
the radii ratio, $R_p/R_s=0.0576\pm 0.0009$, the orbital period, $P=4.8878162\pm 0.0000071$ days,
and the orbital inclination, $i_o=88.5^\circ\pm 0.6^\circ$. 
We fix each parameter above at the central value and use the light curve model by \citet{Ohta2009}.
Assuming a quadratic limb-darkening law, we fix the limb-darkening coefficients 
for the transit light curve as $u_1=0.35$ and $u_2=0.26$ \citep{Claret2004}.
The remaining free parameters are as follows: the midtransit time $T_C$ determined by the simultaneous 
photometry, the RV semi-amplitude $K$, the orbital eccentricity $e$, 
the argument of periastron $\omega$, the spin-orbit misalignment angle $\la$, the projected
stellar rotational velocity $v \sin i_s$, and the RV offset between the Subaru and Keck datasets. 
The $\chi^2$ statistic in this case is expressed as
\begin{eqnarray}
\label{chi2_hat11}
\chi^2 &=& \sum_i \left[ \frac{v^{(1)}_{i,{\rm obs}}-v^{(1)}_{i,{\rm model}}}
{\sigma^{(1)}_{i}} \right]^2+
\sum_j \left[ \frac{v^{(2)}_{j,{\rm obs}}-v^{(2)}_{j,{\rm model}}}
{\sigma^{(2)}_{j}} \right]^2,
\end{eqnarray}
where $v^{(1)}_{i,{\rm obs}}$ and $v^{(1)}_{i,{\rm model}}$ 
are the RV values,
and $v^{(2)}_{j,{\rm obs}}$ and $v^{(2)}_{j,{\rm model}}$
are the relative flux values,
obtained by the observations and model calculations, respectively.
The RV and flux errors are represented by $\sigma^{(1)}_{i}$ and $\sigma^{(2)}_{j}$, respectively.
Note that we adopt a ``stellar jitter" of 
$\sigma_{\mathrm{jitter}} = 4.1$~m~s$^{-1}$ so that the resultant reduced 
$\chi^2$ for the observed RVs becomes unity,
and compute modified RV errors $\sigma^{(1)}_{i}$ by
\begin{eqnarray}
\label{error}
\sigma^{(1)}_{i} &=& \sqrt{\sigma_{0,i}^2+\sigma_{\mathrm{jitter}}^2},
\end{eqnarray}
where $\sigma_{0,i}$ are the reported errors by the RV analysis. 
We used the modified RV errors $\sigma^{(1)}_{i}$ for estimating 
errors of the free parameters.

\section{Results\label{s:sec4} and Discussion}

\begin{figure}[pthb]
 \begin{center}
  \FigureFile(85mm,85mm){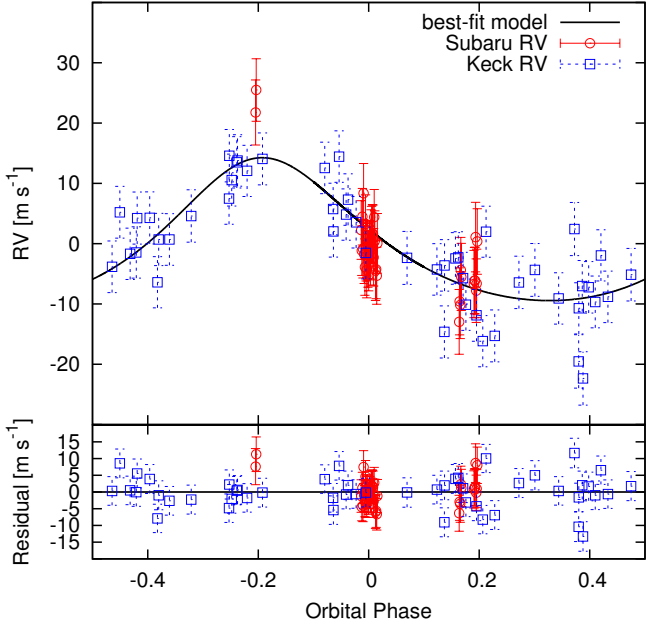} 
 \end{center}
  \caption{RVs taken by the Subaru/HDS (red) and the published 
  ones taken with Keck/HIRES (blue). The best-fit
  curve is shown in the solid (black) line. Each RV error shown in this figure includes stellar jitter. 
  The RV residuals from the best-fit curve are shown in the bottom panel.
  }\label{RV_all}
\end{figure}

\begin{figure}[pthb]
 \begin{center}
  \FigureFile(85mm,85mm){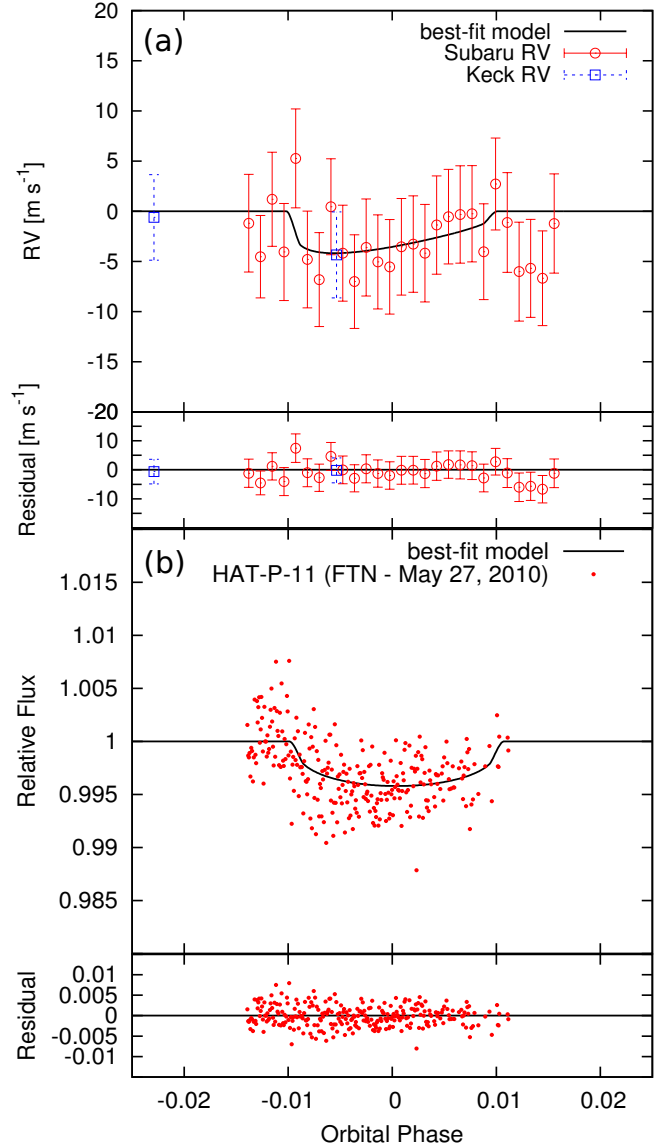}
 \end{center}
  \caption{
  (a) Same as Figure \ref{RV_all}, but zoomed around the transit after subtracting the orbital motion
  (the Keplerian plus constant long-term drift).
  (b) Simultaneous photometry of HAT-P-11 and their residuals from the best-fit light curve, obtained 
  on UT 2010 May 27 with FTN and the Pan-STARRS-Z filter. The best-fit value for the transit center 
  is 2455343.95000 $_{-0.00227} ^{+0.00173}$ [HJD]. 
  Note that we have less photometric data around the transit-end
  due to bad weather conditions. 
  }\label{RV_transit}
\end{figure}

We determine the model parameters so that the $\chi^2$ statistic takes
its minimum value by the AMOEBA algorithm
(see \cite{Narita2008, Narita2010}).
Figure \ref{RV_all} presents the phase-folded RVs obtained by Subaru and Keck,
along with the best-fit RV curve. In this figure,
we subtract the long-term RV variation $\dot{\gamma}$ from the observed values.
In Figure \ref{RV_transit}, we also show the RVs around the transit (phase $\sim 0$), 
as well as the simultaneous photometry with FTN.
The observed RVs show a persistent blue-shift throughout the transit, 
suggesting a spin-orbit misalignment.
The best-fit values for the six parameters are summarized in Table~\ref{hyo2}.
The $1\sigma$ uncertainty for each parameter is estimated by the criterion of $\Delta \chi^2=1.0$.
The spin-orbit misalignment angle $\la$ is estimated as $\la=103_{~-19^\circ}^{\circ+23^\circ}$.
Since the Keck RV dataset covers most of the orbital phase outside of the transit, 
the estimated values for $K$, $e$, and $\omega$ are in good agreement with the reported values
by \citet{Bakos2010}.

\begin{table}[htb]
\caption{The best-fit parameters}\label{hyo2}
\begin{center}
\begin{tabular}{lc}
\hline
Parameter & Best-fit value \\
\hline
$T_C $ [HJD] & 2455343.95000$_{-0.00227} ^{+0.00173}$ \\
$K$ & 11.8 $\pm$ 0.9 [m s$^{-1}$]\\
$e$ & 0.205 $\pm$ 0.036 \\
$\omega$ & 351$_{-12}^{+16}$ [$^\circ$]\\
$\la$ & 103$_{-19}^{+23}$ [$^\circ$]\\
$v\sin i_s$ & 2.09$_{-0.93}^{+0.98}$ [km s$^{-1}$]\\
\hline
\end{tabular}
\end{center}
\end{table}

The estimated stellar rotational velocity
$v\sin i_s = 2.09^{+0.98}_{-0.93}$~km~s$^{-1}$ by our RM analysis agrees with 
the spectroscopically measured value within 1$\sigma$.
In order to confirm the robustness of the estimated spin-orbit misalignment angle $\la$,
we test the following two cases. 
First, instead of letting $v \sin i_s$ be a free parameter,
we fix it as $v\sin i_s=1.5$ km s$^{-1}$ (the spectroscopically measured value) and 
fit the other five parameters.
As a result, we obtain $\la=106_{~-23^\circ}^{\circ+25^\circ}$, in good agreement with the main result described above.
Second, we changed the planet to star radius ratio to the value determined by
\citet{Dittmann2009}, of $R_p/R_s=0.0621\pm0.0011$, based on a photometric follow-up
observation of the system. Adopting their results we obtain $\la=100_{~-21^\circ}^{\circ+24^\circ}$, 
which is also consistent with the main result within  1$\sigma$.
As expected in this case, we obtain a slightly smaller rotational velocity 
($v\sin i_s=1.73_{-0.86}^{+0.83}$ km s$^{-1}$) than the main result.
In both cases above, we find no significant difference in the other fitted parameters ($K,~e,~\omega$) from the main result.

The spin-orbit misalignment angle $\la$ is very sensitive to the RVs taken out-of-transit,
which determine the Keplerian motion of HAT-P-11b.
Our new dataset contains two, nine, and twelve out-of-transit RVs taken
on UT 2010 May 21, 27, and July 1, respectively. 
However, the number of RVs might not be sufficient 
in systems as HAT-P-11, where the planet is small and 
its host star is active.
In addition, since the observations in May and July are 
separated by roughly a month, the long-term
RV drift reported by \citet{Bakos2010} might have affected the result.
If the long-term RV drift is actually caused by a secondary planet lurking in the HAT-P-11 system, the RV drift should modulate with time.
In order to test the effect of a RV drift variation, we artificially increased the RV drift from 
$\dot{\gamma}=0.0297$ m s$^{-1}$ day$^{-1}$ to $\dot{\gamma}=0.10$ m s$^{-1}$ day$^{-1}$.
This treatment resulted in a slightly larger value for the spin-orbit misalignment angle 
($\la=109_{~-24^\circ}^{\circ+32^\circ}$) and a smaller value for 
the projected stellar rotational velocity 
($v\sin i_s=1.68_{-0.94}^{+1.12}$ km s$^{-1}$), both of which are consistent with the main result
shown in Table~\ref{hyo2}.  
This result indicates that the long-term RV drift has less impact on 
estimating $\la$, as long as $\dot{\gamma}$ does not exceed 0.10 m s$^{-1}$ day$^{-1}$.

Although the measurement of the RM effect in the HAT-P-11 system seems challenging
due to the small size of the planet and large stellar jitter, our result
suggests a significant spin-orbit misalignment of the system. 
As we have described in Section \ref{s:sec1}, five out of the seven eccentric
hot-Jupiters where the RM effect has been observed have significant
spin-orbit misalignment.
Our result suggests that the super-Neptune HAT-P-11b migrated to its present
location by dynamical scattering or a long-term perturbation by
an outer body, similarly to other eccentric hot-Jupiters.
By contrast, the fraction of misaligned systems with circular orbits is significantly smaller
(e.g., CoRoT-1, HAT-P-7, WASP-15; \cite{Pont2010, Narita2009b, Winn2009b, Triaud2010}).

\citet{Winn2010} suggested the interesting possibility that hot-Jupiters have large initial spin-orbit
misalignment caused by dynamical processes, but the host stars' obliquity could \textit{decline}
due to tidal interactions between hot-Jupiters and the stellar convective zone.
Since convective zones are particularly well-developed in cooler and  less massive stars, we are likely
to observe spin-orbit alignment around cool host stars.
Although HAT-P-11 is a very cool star ($T_\mathrm{eff}=4780\pm 50$ K),
this hypothesis also claims that the decay timescale of host star's obliquity is
larger when the planet has a lower mass and a distant orbit from the host star (see eq.~2 of \cite{Winn2010}).
According to their criteria, the HAT-P-11 system can show a spin-orbit misalignment
mainly because of the lower mass of HAT-P-11b.

Comparison of the spin-orbit misalignment angle for HAT-P-11b with those of other 
transiting hot-Neptunes (e.g., GJ~436b and Kepler-4b) is quite interesting,
because their host stars are of different spectral types (Kepler-4 is
a G0 star and GJ 436 is an M2.5 star), while HAT-P-11b, GJ 436b,
and Kepler-4b have similar planetary radii and masses
(e.g., Butler et al. 2004; Shporer 2009; Borucki et al. 2010).
Further observational investigations of
formation and migration history of hot-Neptunes will be an interesting
topic in the next decade.

\section{Summary\label{s:sec5}}
We have measured the RM effect for one of the smallest transiting exoplanets known to date, HAT-P-11b. The exoplanet has a significant eccentricity and is an interesting target for the RM effect.
The observed RV anomaly during the transit suggests a significant 
spin-orbit misalignment of the system, of $\la= 103_{~-19^\circ}^{\circ+23^\circ}$.
To confirm the spin-orbit misalignment decisively, however, it is necessary to measure
even more RVs of the system during and outside transits,
as well as to better characterize the long-term RV variation.
Although challenging, the measurement of the RM effect for Neptune-sized exoplanets will extend
the parameter space where planetary formation and migration theories will be studied
in the near future.
\\

We are very grateful to Joshua N. Winn and Atsushi Taruya for helpful comments on
our results.
This paper is based on data collected at Subaru Telescope,
which is operated by the National Astronomical Observatory of Japan.
We acknowledge the support for our Subaru HDS observations
by Akito Tajitsu, a support scientist for the Subaru HDS.
This paper also uses observations obtained with facilities of the 
Las Cumbres Observatory Global Telescope.
The data analysis was in part carried out on common use data analysis
computer system at the Astronomy Data Center, ADC,
of the National Astronomical Observatory of Japan.
T.H. and N.N. are supported by Japan Society for Promotion of Science
(JSPS) Fellowship for Research (DC1: 22-5935, PD: 20-8141).
M.T. is supported by the Ministry of Education, Science,
Sports and Culture, Grant-in-Aid for
Specially Promoted Research, 22000005.
We wish to acknowledge the very significant cultural role
and reverence that the summit of Mauna Kea has always had within
the indigenous people in Hawai'i.



\end{document}